\documentclass{optica-article}

\journal{opticajournal} 

\articletype{Research Article}

\usepackage[utf8]{inputenc}

\usepackage{lineno}

\begin{document}

\title{Compact Cavity-Enhanced Aerosol Detector using Incoherent Light Sources}

\author{Jacob Williamson,\authormark{1,*} Pranav Chamakkad Muthukrishnan,\authormark{1} Srushti Nandanwar,\authormark{1} Shuaifeng Guo,\authormark{1} and Chandra Raman\authormark{1}}

\address{\authormark{1}School of Physics, College of Sciences, Georgia Institute of Technology, 837 State St NW, Atlanta, GA 30332}

\email{\authormark{*}jwilliamson67@gatech.edu} 


\begin{abstract*} 
We have realized a compact optical particle counter utilizing  enhancement of light scattering within a high finesse Fabry-Perot optical cavity.  In contrast to laser-based approaches such as cavity ringdown spectroscopy we use the light stream from both superluminescent and light-emitting diodes that have no longitudinal coherence.  This eliminates the vibration sensitivity that is typical of laser-based cavity methods.  The use of the transmission mode of detection allows us to reduce the cavity mirror separation to below 1 cm, with no obvious limit to miniaturization.  Typical light scattering instruments are larger, in part due to their sensitivity to background signals from the light source.  Our approach paves the way toward a new generation of compact and portable instruments.  
A simultaneous comparison of the scattering signals with a commercial particle counter shows that our cavity is also sensitive to ultrafine particles below 300~nm diameter that are typically not recorded in such counters.
\end{abstract*}

\section{Introduction}

Small airborne particles whose diameter is below 10 micrometers are associated with negative health effects and are important to quantify in both outdoor and built environments \cite{us_epa_particulate_2016,cohen_estimates_2017, bench_measurement_2004, paisi_modeling_2023, sharma_four_2022, gu_review_2020, contini_contribution_2021, sun_exposure_2019, pope_iii_health_2006,el-sharkawy_indoor_2014}.  In addition, ultrafine particles (below 100~nm in diameter) are of increased interest in recent years due to their ubiquity and a lack of understanding of their effect on human health \cite{wright_small_2019, kwon_ultrafine_2020, nazaroff_ten_2023}.  Light scattering is an old and well-established technique for detection of such particles, with both precision instruments and low-cost sensors available commercially \cite{kerker_light_1997, walser_parametrization_2017, liu_response_1992}.  In spite of their ubiquity, these sensors are far bulkier and more difficult to miniaturize in comparison to modern sensors for detection of other harmful compounds in the built environment, for example, VOCs \cite{edwards_compact_2012, mohammadzadeh_unique_2023} and CO$_2$ \cite{ledermann_piezoelectric_2004, ng_miniaturized_2022} detectors.  Additionally, while CO$_2$ is a large component of air quality, CO$_2$ measurements alone have been shown to be insufficient to fully assess indoor air quality \cite{westgate_using_2022}, and so particle detection is necessary to fully characterize air quality.  

A key issue for particle detector miniaturization is fundamental to the physics of light scattering from bright laser beams that maximize the scattering signal: the beams also produce large background scattering from nearby surfaces which must be mitigated via beam dumps.  If the detector could be placed directly in the beam path to measure the change in optical transmission the sensor could be miniaturized considerably.  However, the relative change in transmission is the ratio of particle scattering cross-section to the incident beam area, and thus is usually quite small.  For example, a 0.5 micron diameter particle in a Gaussian beam of 50 micron radius results in a fractional change of $5\times 10^{-5}$, which is very challenging to measure.

In an attempt to solve these problems, cavity-enhanced methods of aerosol detection have been used in the past. Both coherent light sources \cite{schuster_detection_1972, liu_measurement_2013, butler_optical-feedback_2009, butler_cavity_2007} as well as incoherent ones have been used.  The latter include incoherent broad-band cavity-enhanced absorption spectroscopy (IBBCEAS) with either an LED  \cite{zhao_development_2014,zhao_development_2017, stollberger_direct_2023, bahrini_incoherent_2018,varma_light_2013} or a supercontinuum source \cite{saseendran_dual-cavity_2020, wang_novel_2023}, as well as cavity attenuated phase shift (CAPS) methods based on detectors that use a pulsed light source and a photodiode \cite{massoli_aerosol_2010,yu_direct_2011, kebabian_optical_2007}.  However, these detectors are large, with some instruments being 70~cm \cite{zhao_development_2017}, or even above 90~cm \cite{saseendran_dual-cavity_2020} in length.  While increasing the cavity path length improves the sensitivity to intracavity absorption, it is strictly unnecessary for detecting the passage of a single aerosol whose diameter lies below 10 microns, and renders the device less portable. 

In this work, we utilize cavity enhancement to efficiently detect the change in transmitted light from the passage of microscopic aerosol particles.  In contrast to previous work, we use a much shorter cavity with a separation of only 8~mm and a physical volume $\lessapprox1$~cm$^3$.  While direct measurement of scattered light has been the tool of choice, a cavity transmission measurement scheme has significant advantages.  For one, it is sensitive to the {\em entire} electromagnetic cross-section, including both absorption and the total elastic scattering from the particle, whereas light scattering only measures the {\em differential} cross-section.  Moreover, light scattering measurements typically place the detector at 90 degrees to avoid contamination from the bright light source.  For Mie scattering, much of the light is scattered at small angles from the incident beam direction and thus is not detected by conventional scattering detectors.  However, any amount of scattering, including forward and back scattering, will remove light from the cavity mode in our detector and thus cause a drop in the transmission.  A second and key advantage is that the transmission mode is not sensitive to background scattering and thus poses no limit to further sensor miniaturization.  

In a traditional scattering detector, the photons from a beam have a single chance to interact with a particle in the beam.  However, with a cavity-based detector, each photon in the beam has multiple chances to interact with a particle.  The number of chances is related to the cavity finesse
\begin{equation}
    \mathcal{F} = \frac{\pi\sqrt{R}}{1-R}=\pi\cdot n_\text{bounces}
\end{equation}
where $R$ is the cavity mirror reflectivity and $n_\text{bounces}$ is the average number of mirror bounces a photon experiences in the cavity.  Between each bounce the photon has one chance to interact with an aerosol particle, so the cavity enhancement factor is  $\mathcal{F}/\pi$.  Improvements in the signal (and therefore the sensitivity) can be achieved by increasing the mirror reflectivity/cavity finesse.  In this work, we demonstrate a sensitivity similar to a commercial particle counter when using a cavity whose finesse in the range of a few 100, and even greater improvements in sensitivity \textit{and} size can be obtained in a straightforward manner by improving the finesse, since smaller cavities can have longer effective path lengths with greater finesse. 




\section{Materials and Methods}

\subsection{Experimental Setup}

\begin{figure}[ht!]
\centering\includegraphics[width=\textwidth]{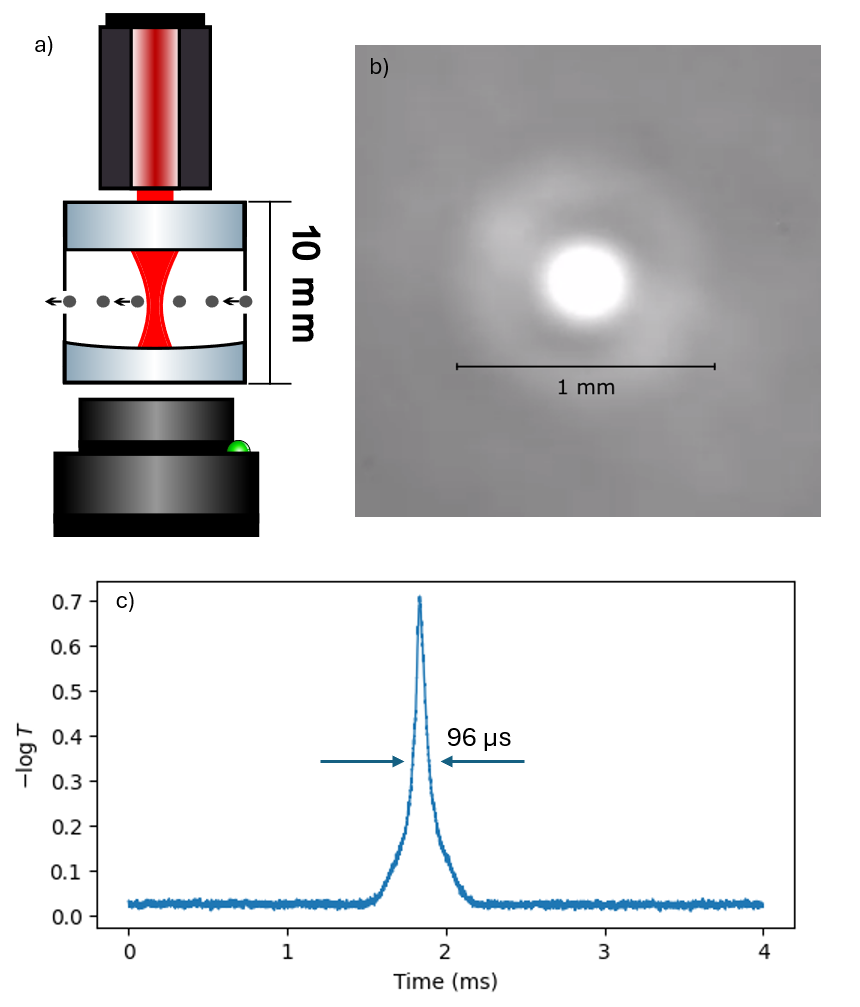}
\caption{Summary Image.  Image a) shows a cartoon schematic of the operating principle of the optical cavity particle counter.  A laser is shined into a Fabry-Perot cavity while the transmission is monitored either by a camera or photodiode, as seen in b).  Particles are passed through the cavity via an airflow and the change in transmission is then measured, as shown in c).}
\label{fig:one}
\end{figure}


Figure \ref{fig:one}a) shows the operating principle of our optical cavity sensor.  It is based on the change of transmission in a Fabry-Perot resonator caused by a particle entering the cavity mode.  The light source, in our case either a superluminescent diode (SLD) or LED, is coupled to the cavity modes and the resulting transmission is detected by a photodiode or camera on the opposite end.  Typically, a Fabry-Perot cavity is used either to lock a laser to a specific frequency or to measure the frequency of a laser.  In contrast, we are using a broadband source instead of a laser, which excites multiple modes across multiple free spectral ranges (FSRs) so instabilities in both the cavity and the light source are averaged out.  Additionally, single-frequency cavities are highly sensitive to vibrations, since any change in the cavity length will change the resonance frequencies; this is exacerbated with high finesse cavities as higher finesse means a smaller resonance peak width and so a more sensitive cavity to small deviations.  Therefore, by using a broadband source, we are also able to stabilize the cavity against vibrations.  On the downside, the absolute power output of the cavity is lower by a factor of the finesse, but the cavity sensitivity to particle transits is nonetheless increased (see Supplemental Information).  

As can be seen in image b) of Figure \ref{fig:one}, the majority of the light in the cavity is coupled to the central TEM$_{00}$ modes.  However, there is also a dim ring around the center which is indicative of some higher order modes.  This is what causes a small amount of broadening seen on the bottom of the peak in image c): particles transiting across the cavity will first go through the higher order modes, which have proportionally less light and so will cause a smaller change in transmission than when the particle transits the main mode.

\begin{figure}[ht!]
\centering\includegraphics[width=\textwidth]{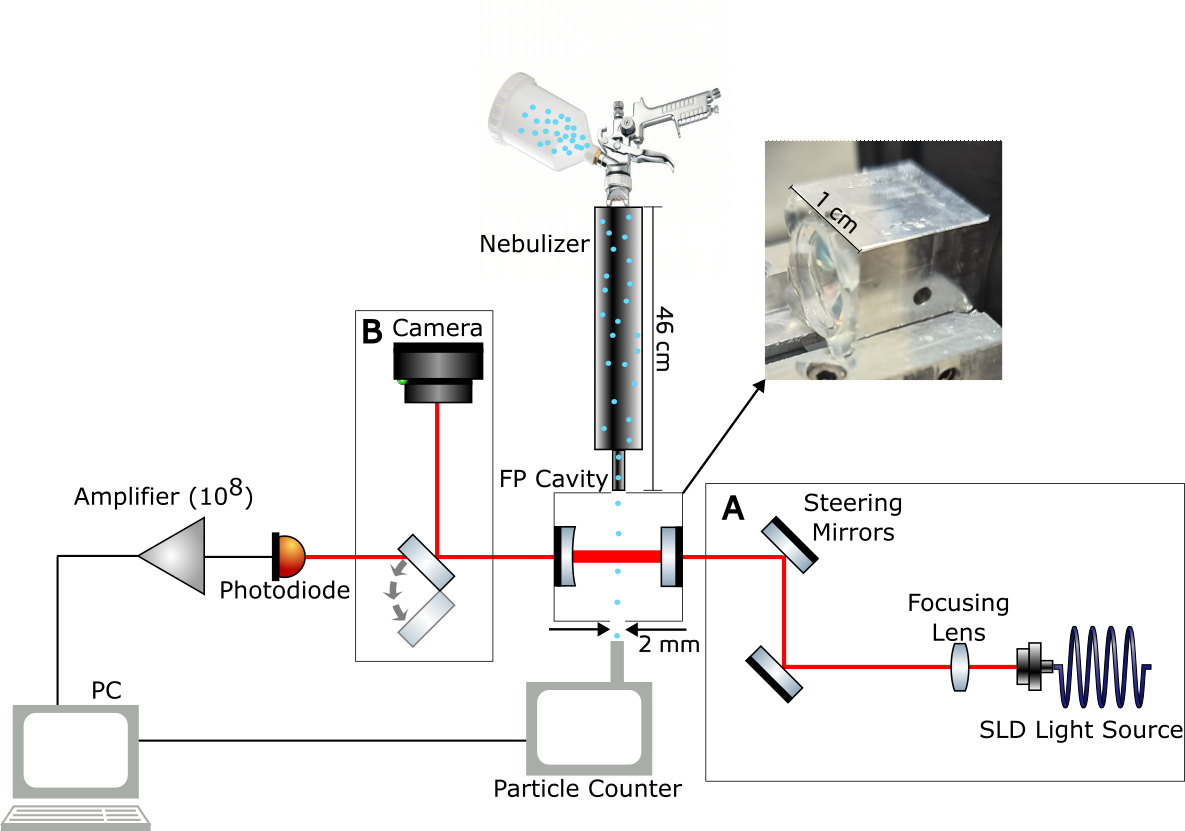}
\caption{Setup for particle detection.  For the main experiment, a superluminescent diode (SLD) is sent through a focusing lens mode-matched to the cavity and steered using the steering mirrors.  After going through the cavity, the transmission is either measured via a photodiode or sent via a flip mirror to a camera to monitor the transmission mode.  Inset image shows the cavity housing used in the SLD experiment.  In the secondary experiment, all of the optics in box A are replaced with a single LED placed close to the cavity and the optics in box B are removed while the photodiode is moved close to the cavity.  Sprayer image from \cite{granger_8_nodate}.}
\label{fig:setup}
\end{figure}

Figure \ref{fig:setup} shows a schematic of our setup for particle detection and counting.  Light from an SLD centered at 810~nm (ThorLabs SLD810S) is piped in via an optical fiber and focused to match to the fundamental TEM$_{00}$~mode of the Fabry-Perot cavity using the focusing lens.  The cavity itself is housed in a custom-machined aluminum housing, approximately 1~cm$^3$ in volume, with a 2~mm diameter through hole to allow aerosols to pass through the cavity (see inset).  The light beam is delivered into the cavity using the steering mirrors.  Particles of aerosolized isopropyl alcohol are injected via a SPEEDAIRE spray gun (Model 48PX88) powered by compressed air at 50 psi.  The spray passes through a nebulizer to allow for the particle distribution to evolve.  This is similar to the processes used by condensation particle counters to detect particles below 300~nm in size.  The particles then enter the cavity housing through the hole in the cavity housing, followed by a Particles Plus Model 5306 Particle Counter to be counted at nearly the same time ($\lessapprox 1$~s of lag time).  The particle counter also acts as an air pump for the cavity and pulls the air through at a rate of 0.1 CFM (2.83 LPM).  

Using the flip mirror in box B, we can either check the alignment directly using a camera (see Figure \ref{fig:one}b), or measure the transmission of the cavity using a photodiode with a transimpedance amplifier (FEMTO Model DLPCA-200) set at $10^8$ gain.  The photodiode signal is measured using a PicoScope Model 5242D and the results of the particle counter are measured simultaneously using the PC.  An example of a particle transit as seen by the photodiode is shown in Figure \ref{fig:one}c).  For tests with an LED instead of an SLD, we do not use mode-matching as the angular distribution of the light makes precise mode matching impossible.  Because of this, the LED and detector must both be placed as close as possible.  Therefore, the entirety of box A in Figure \ref{fig:setup} is replaced with an LED (ThorLabs LED430L) centered at 430~nm placed as close to the right-hand cavity mirror as possible.  Box B is also removed and the photodiode is placed as close as possible to the left-hand cavity mirror to maximize detection.  


\subsection{Scattering theory of small particles in an optical cavity}

In order to compare the transmission measurements from our cavity with the particle counts given by the particle counter, we must derive the effect of a particle on the transmission of a cavity in a broadband context.  The interaction between a particle and the cavity mode may be treated using standard tools of electromagnetic scattering theory \cite{zangwill_modern_2012, mishchenko_scattering_2002, hulst1981light}.  This in general includes elastic scattering and absorption, both of which cause a reduction in the light transmitted through the cavity.  To compute the change in light level caused by the particle's transit, we must first calculate the effect of the particle on a single 'pass' of the cavity mode (see Figure \ref{fig:one}).  This field is given by
\begin{equation}
    \vec{E} = \vec{E}_{inc} + \vec{E}_{sc}
    \label{eq:one}
\end{equation}
which is the sum of the incident electric field on the particle $\vec{E}_{inc}$ and the forward scattered field by the same particle, $\vec{E}_{sc}$.  Provided that the scattered field is much weaker than the incident field (usually the case), the single-pass electric field transmission coefficient is
\begin{equation}
    \alpha = \frac{E}{E_{inc}} = 1 + i A 
    \label{eq:two}
\end{equation}
where $A$ is proportional to the forward scattering amplitude $f(0)$ which is calculated from Mie scattering theory.  The real and imaginary parts of $f(0)$ lead to the intra-cavity phase shift and reduction of the cavity mode energy, respectively.  

As we show in the Supplement, the single pass intensity transmission coefficient is 
\begin{equation}
    |\alpha|^2 = e^{-\kappa} = {\rm exp}\left({-\frac{\sigma }{A_b}}r_0(t) \right) 
    \label{eq:alpha2}
\end{equation}
where the single-pass extinction coefficient $\kappa = \frac{\sigma}{A_b}r_0(t)$
and where $\sigma,A_b$ are the total electromagnetic cross-section for the particle and the intracavity beam area, respectively.  
The factor $r_0(t) = I({\bf r}(t))/I(0)$ accounts for the particle's trajectory through the spatially varying beam profile.


We can then use $|\alpha|^2$ to compute the cavity transmission coefficient using the standard formulas for a Fabry-Perot cavity.  The transmission is given by
\begin{equation}
    T = \frac{(1-R)^2|\alpha|^2}{1+R^2|\alpha|^4-2R|\alpha|^2\cos(\phi)}
\end{equation}
where $R$ is the reflectivity of the cavity mirrors and $\phi=2\pi\frac{L}{\lambda}$ is the round trip phase shift.  Since we are using a broadband source, we need to integrate this equation over the range of wavelengths.
However, for our cavity the free spectral range (FSR) is $\frac{\lambda^2}{2L}\approx.04$~nm (19.7~GHz), and since our SLD has a bandwidth of approximately 32.5~nm (16~THz), it covers many FSRs.  Therefore, an accurate estimate of the cavity transmission is made by simply summing over one free-spectral range, i.e., integating Eqn.\ \ref{eq:transmission} from $\phi = 0$ to $2\pi$:
\begin{equation}
    T=\int_{0}^{2\pi}\frac{(1-R)^2|\alpha|^2}{1+R^2|\alpha|^4-2R|\alpha|^2\cos(\phi)}d\phi
    \label{eq:transmission}
\end{equation}
This formula is the central one used to convert measured transmission to particle size in our data sets.

\section{Results and Discussion}

\begin{figure}[ht!]
\centering\includegraphics[width=\textwidth]{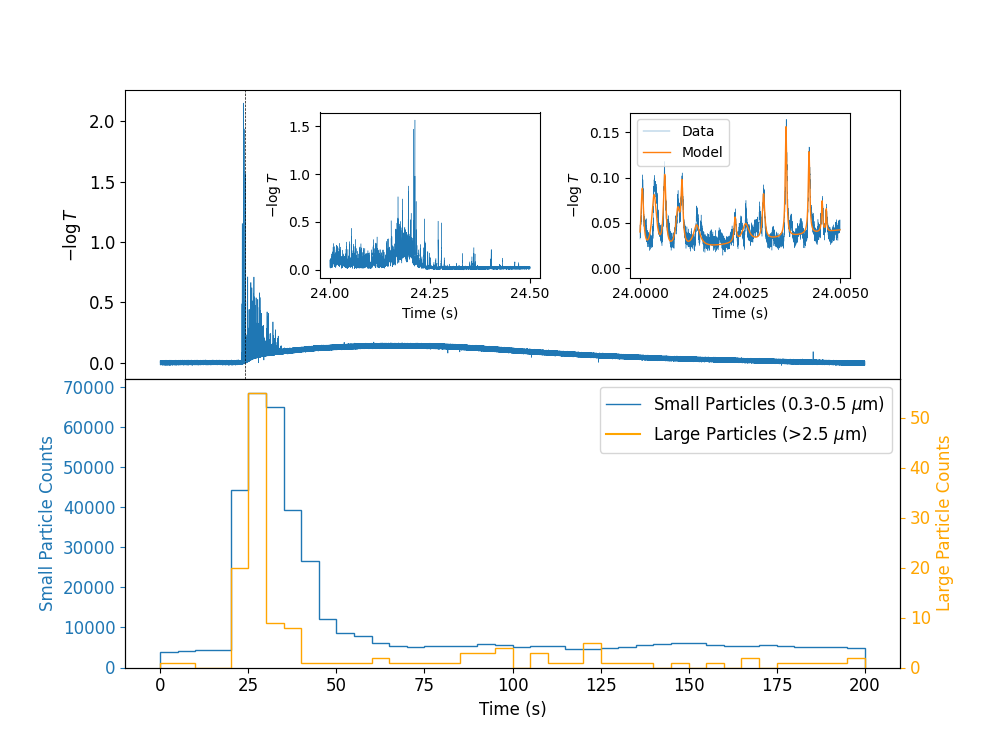}
\caption{The data from the SLD experiment.  The top image shows the photodiode trace, in the form of –log(T).  The inset is a close-up of 5 milliseconds to show an example of the peak fitting process.  The bottom image has both the large particle counts and small particle counts from the particle counter to compare with the cavity data.}
\label{fig:SLDRaw}
\end{figure}

Figure \ref{fig:SLDRaw} shows the raw cavity transmission data and compares it with both small and large raw particle counts from the particle counter.  The particle counter measures across a 5-second interval and reports the counts detected in that time, which are then plotted in the bottom plot as two histograms.  The large particle count histogram is composed of a sum of size bins above 2.5 $\mu$m.  In general, we observe that particles create a set of peaks in the transmission data between 20~s and 35~s, which corresponds to a peak in particles detected by the particle counter.  However, looking at the particles in the range of 300-500~nm, the peaks are correlated, but there is a longer tail in the particle counts which does not correspond with the peaks seen in the transmission data.  When examined more closely at millisecond timescales and below (see insets) the peaks  clearly correspond to individual particle transits across the cavity mode (see also Figure \ref{fig:one}c).  Additionally, we observe a background 'shoulder' which does not follow the trend of the particle counts at all.

To better compare the particle counter data with the cavity data, we extract the peaks from the data using a piece-wise fitting method by means of the \verb|lmfit| Python package \cite{newville_lmfitlmfit-py_2024}.  We split the data into 10~ms 'chunks', count the number of peaks $N_i$ above a certain peak prominence value, then fit the chunk to a line plus $N_i$ Lorentzian peaks, where $i$ is an index ranging from $1$ to the number of 10~ms chunks in the data $N_\text{chunks}$ (in this case, $N_\text{chunks}=20000$).  Because there are thousands of individual transits with varying widths, ranging from about 20~$\mu$s to around 200~$\mu$s, splitting the data in this way minimizes the computing time without overly compromising the peak extraction; splitting into shorter chunks leads to excessive loss of peak data and longer chunks exponentially increase the computation time.  Additionally, because of the shoulder visible in \ref{fig:SLDRaw}, attempting to fit the entire data set would require \textit{a priori} knowledge of the functional form of that shoulder.  To pick out the peaks, we use the \verb|find_peaks| function in the \verb|scipy.signal| package \cite{virtanen_scipy_2020}.  We tested several prominence values until we found a peak prominence high enough to prevent detection of peaks which may be due to noise but also low enough to maximize the number of actual peaks fit.  

The functional form of the $i$-th chunk fit is:
\begin{equation}\label{eq:fittingeqn}
f_i(t_i)=m_it_i+b_i+\sum_{n=1}^{N_i}\frac{A^{(i)}_n}{\pi}\left[\frac{\sigma^{(i)}_n}{(t_i-\mu^{(i)}_n)^2+(\sigma^{(i)}_n)^2}\right]
\end{equation}
where $m_i,b_i$ are the slope and y-intercept of the linear portion of the $i$-th chunk; $A^{(i)}_n, \sigma^{(i)}_n,\mu^{(i)}_n$ are the amplitude, half width at half maximum (HWHM), and center of the $n$-th Lorentzian in the $i$-th chunk; $N_i$ is the total number of peaks in the $i$-th chunk (if $N_i=0$ then only a linear fit is performed); $(i-1)*T\leq t_i\leq i*T$ is the time range of the chunk in milliseconds; and $T$ is the duration of each chunk (in this case, 10~ms).  To improve fitting times and minimize fit errors, the center of each peak is fixed at the value given by \verb|find_peaks|, so only the amplitude and HWHM are fit.  These parameters are also constrained to be positive to prevent errors in fitting.  An example of one of these fit results is shown in the second inset of Figure \ref{fig:SLDRaw}.  

\begin{figure}[htbp]
\centering\includegraphics[width=\textwidth]{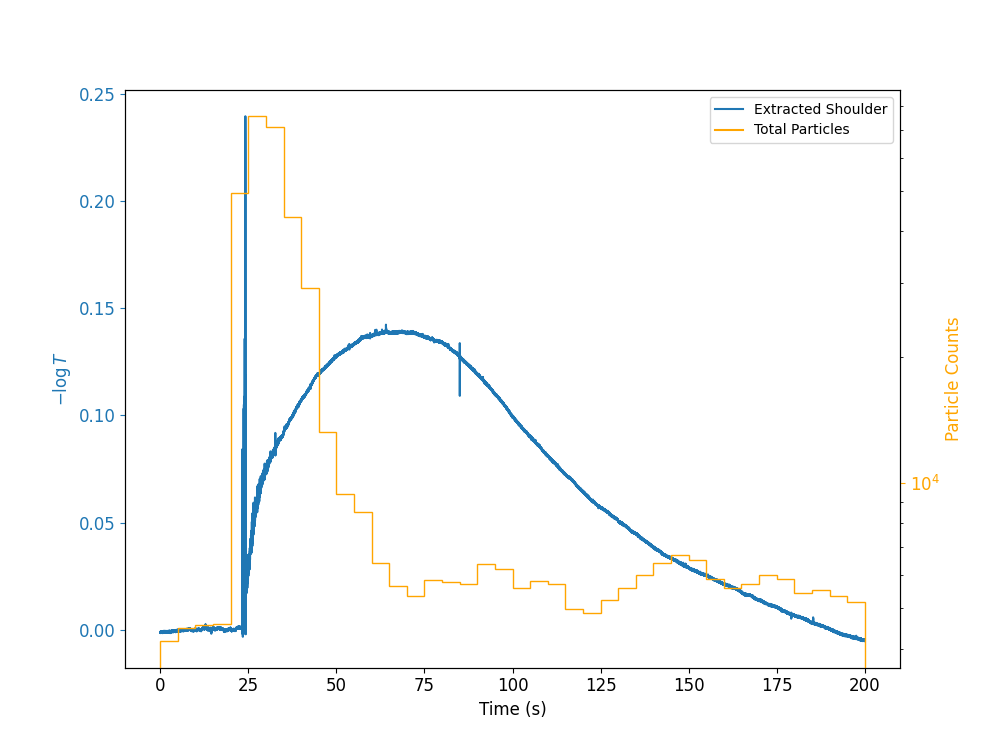}
\caption{The extracted shoulder compared to the particle counter data.  The extracted shoulder is made up of piece-wise linear sections 10 milliseconds in length.  The key takeaway is the large hump in the center which is not correlated with the particle counts given by the counter.  This suggests that something other than particles above 300~nm in size is causing this shoulder.}
\label{fig:backparticles}
\end{figure}

Figure \ref{fig:backparticles} shows the extracted shoulder using the linear portions of Equation \ref{eq:fittingeqn} as fit to the raw data in Figure \ref{fig:SLDRaw}, as in the function
\begin{equation}
    l(t)=\sum_{i=1}^{N_{\text{chunks}}}(m_it+b_i)*\text{rect}\left(\frac{t}{T}-i+\frac{1}{2}\right)
\end{equation}
where rect$(t/a)=\Theta(t+a/2)-\Theta(t-a/2)$ is the rectangular function (also known as the Heaviside Pi function), $m_i,b_i$ are the slope and y-intercept of the linear portion of the $i$-th chunk, $N_{\text{chunks}}$ is the total number of chunks in a single data set (equal to the total measurement time divided by the chunk time of 10~ms), $T$ is the duration of each chunk, and $t$ is in milliseconds. By only plotting the linear portions, the shoulder is more plainly visible.  The main feature we observe is the central shoulder that does not correlate with the particle counts as detected by the Particles Plus counter.  Since a particle larger than 25~$\mu$m (the maximum size registered by the Particles Plus detector) would cause a sharp peak in transmission, we believe this shoulder is due to a cloud of particles below 300~nm in diameter, as such particles are not detectable by the Particles Plus detector.  Further work may shed more light on the underlying mechanism causing this feature.

One potential drawback of this fitting and extraction procedure can be inferred from the first inset of Figure \ref{fig:SLDRaw}: If there are enough peaks near enough to each other, the background line is almost completely occluded.  This results in under-estimations of peak sizes as well as inaccurate background estimates, which shows up as a "spike" in the extracted shoulder in Figure \ref{fig:backparticles}.  Attempts to minimize the spike shown in Figure \ref{fig:backparticles} required increasing the chunk time from 10~ms and resulted in exponentially longer computation times with minimal reduction in feature size.  Further exploration can be found in the Supplemental Information document.

\begin{figure}[ht!]
\centering\includegraphics[width=\textwidth]{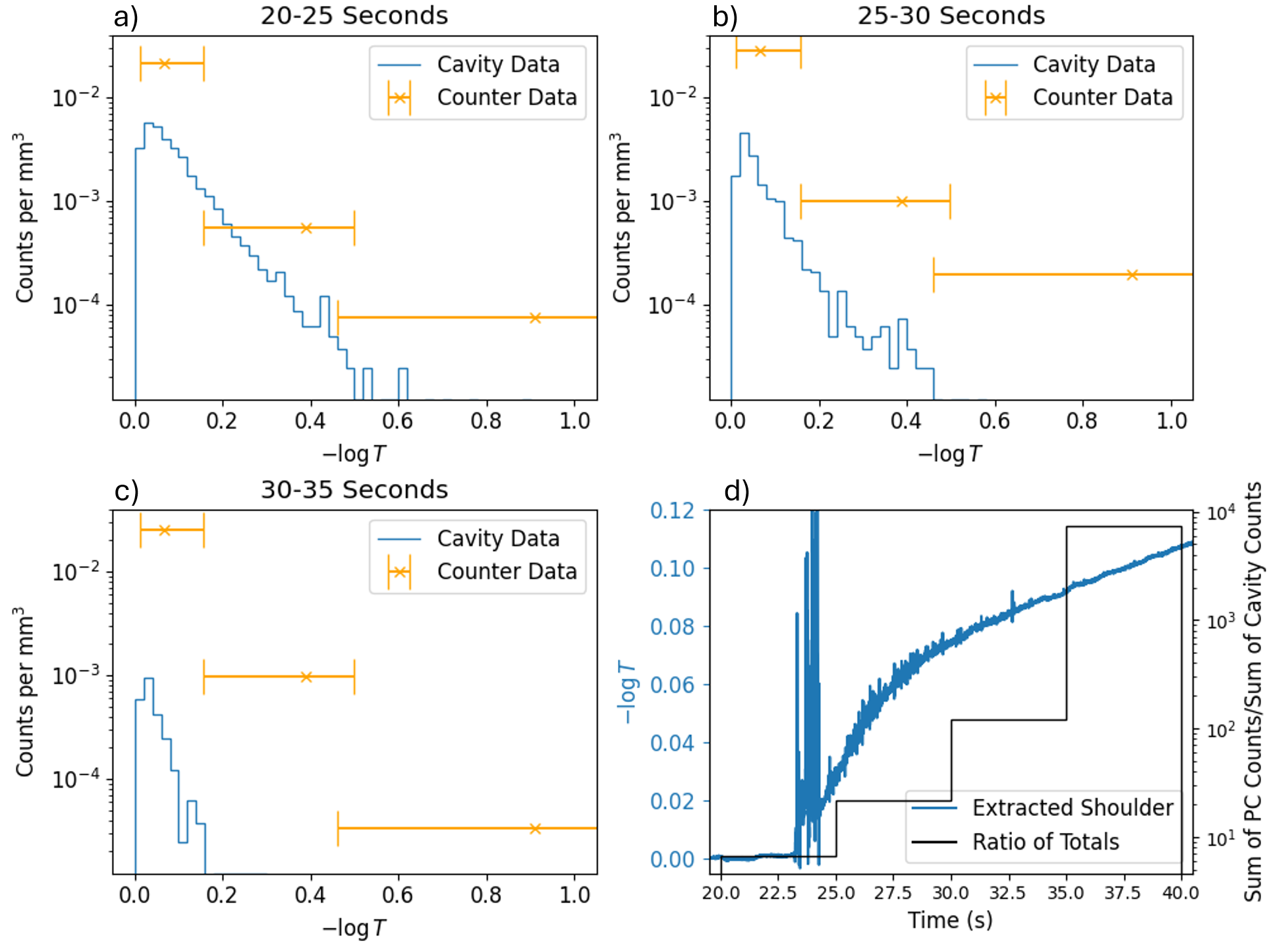}
\caption{Peak counts compared to particle counts for multiple 5-second chunks.  The particle counter only measures in 5-second increments, so the peaks in the trace from Figure \ref{fig:SLDRaw} are fit and binned in each 5 second chunk.  The counter data is placed on the –log(T) axis by calculating the Mie scattering cross-section for each particle size and calculating the resulting transmission for the particle size bins.  A ratio is then taken of the total number of counts and peaks and compared to the relevant background region.  This shows again that the background is not caused by the same particles detected by the particle counter.}
\label{fig:peaksparticles}
\end{figure}

To compare cavity and counter approaches to determining particle sizing, we have plotted a histogram of the extracted peak heights with particle counter data in the form of $-\log T$ in Figure \ref{fig:peaksparticles}.  To make the comparison, the x-axis of the particle counter data is first converted to a value of $-\log T$ by calculating the transmission through a Fabry-Perot cavity for a range of particle sizes in each size bin according to Equation \ref{eq:transmission}.  The three visible points for the particle counter represent, from left to right, the $0.5-1$~$\mu$m, $1-2.5$~$\mu$m, and $2.5-5$~$\mu$m size bins, respectively.  The error-bars on the figure represent the range of transmissions for each size bin as calculated by Equation \ref{eq:transmission}.  Notice there is some overlap between some adjacent size bins: As the particles are all in the regime of Mie scattering, some overlap between size bins is expected due to the shape of the Mie scattering cross-section curve.  Additionally, we assume that each peak is a single particle transit and then use the calculated mode waist and dimensions of the cavity enclosure to calculate the total mode volume through which the particles pass.  

We observe that the general trends of the peak height histograms and the particle counter points are similar, which lends confidence to our data analysis procedure.  However, one can see that the efficiency of detection by the cavity appears to decrease with time, as evident from the growing difference in the absolute number of counts between these two methods.  We believe this change may be due to differences in the dynamics of particle motion within the cavity and the counter due to their being installed in-line with one another, and is not due to any fundamental property of the cavity.  For instance, the cavity mode waist is $55$~$\mu$m while the hole in the cavity enclosure is 2~mm in diameter (see Figure \ref{fig:setup}), so only a fraction of particles moving through the particle counter will pass through the cavity mode, and this fraction may not be constant with time.  Since the particles are injected via a spray, this could also introduce excessive turbulence near the spray time, causing particles to move more erratically and thus be more likely to enter the cavity mode.  Future experiments could mitigate this effect by using two independent airflow setups.  Figure \ref{fig:peaksparticles}d shows that this effect cannot explain the shoulder in the data mentioned earlier.  It compares the ratio of particle counter counts and peak counts to the shoulder $l(t)$ from Figure \ref{fig:backparticles}.  The trend is similar, but not enough to explain the shoulder completely, which is another reason we think the shoulder is detecting ultrafine particles.

\begin{figure}[ht!]
\centering\includegraphics[width=\textwidth]{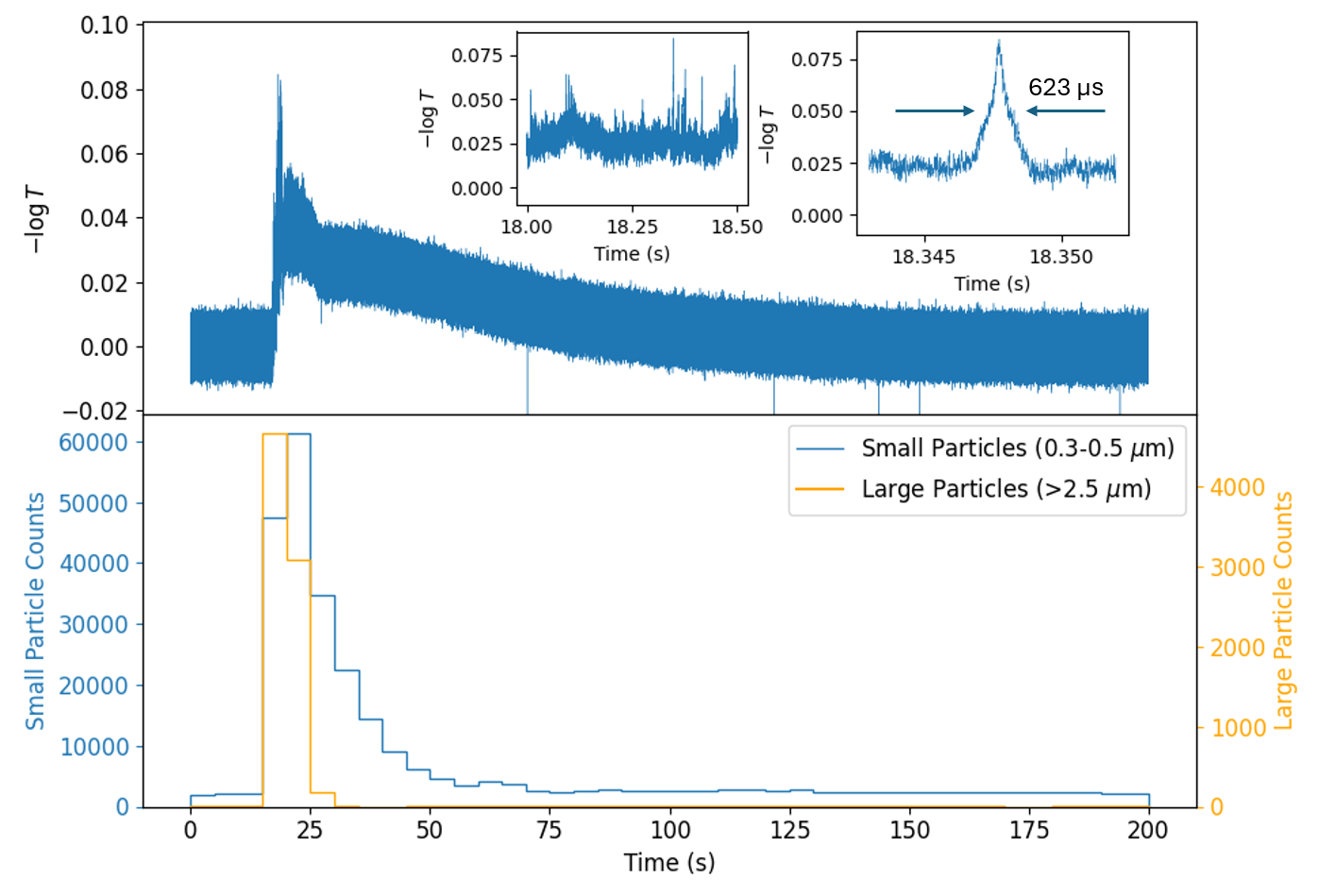}
\caption{The same as Figure \ref{fig:SLDRaw} but from the LED experiment.  A similar shoulder is visible here as well as some similar peaks, though they are much smaller and wider.}
\label{fig:LEDRaw}
\end{figure}

To further explore the possibilities of incoherent broadband light in cavities, make the experiment more compact, and pursue low-cost options, we replace A and B in Figure \ref{fig:setup} with an LED centered at 430-nm and a photodiode, respectively.  Figure \ref{fig:LEDRaw} shows raw data from this LED version of the experiment as compared to particle counter data taken simultaneously.  We observe a series of peaks around the 15-30~s range of the data which corresponds with the peaks in particle counts below.  Additionally, we see another shoulder as we saw in Figure \ref{fig:SLDRaw}.  While this shoulder appears to follow the small particles histogram more closely than with the SLD data, it is still not close enough to fully explain it.  We therefore conjecture that this shoulder is again due to ultrafine particles as with the SLD experiment.

The peaks in the LED transmission data are significantly smaller and wider than the peaks in Figure \ref{fig:SLDRaw}.  This is because the LED setup forgoes mode-matching as it is a diffuse light source.  In our case we have a short plano-concave cavity with stability parameter $g_1 = \infty$ and $g_2 = 1-d/R$, where $d = 8$ mm and $R = 24$ mm are the 2nd mirror's separation from the flat mirror and its radius of curvature, respectively.  For this configuration, we used a ray matrix approach to estimate the intracavity mode.  We assumed, as per manufacturer specifications \cite{thorlabs_thorlabs_nodate}, that the input LED source had position and angular widths of 2 mm and 15 degrees, respectively (root-mean-square values).  This resulted in an intracavity beam width of 2.5 mm rms.  This is in reasonable agreement with the 3 mm FWHM measured width using a razor blade method as detailed in the Supplement.

By comparing the inset in Figure \ref{fig:LEDRaw} with the peak shown in Figure \ref{fig:one}c), we can see that the peak width is increased by approximately 6.5 times.  The larger mode partially explains the longer transit time, but not completely.  If the larger mode was the only factor, we would expect the peak width to increase by approximately 45 times, since the waist of the SLD cavity is about 55~$\mu$m.  Additionally, from Figure \ref{fig:SLDRaw}, we see that the maximum value of $-\log T$ for the SLD experiment is more than 2, while for the LED experiment the maximum value is between 0.08 and 0.09 (approximately a factor of 25 smaller).  This is also partially explained by the lack of mode matching as the LED should couple into all modes equally, and the fractional size difference between the effective cavity mode and the particle scattering cross-section is much smaller than in the SLD experiment.  This also means that the LED-based cavity is more susceptible to noise than the SLD version and requires more noise reduction.  Additionally, the wider peaks cause more overlapping near the spray times, and so the chunk splitting procedure for fitting we used previously with the SLD-based cavity is more difficult.  Therefore, in order to perform similar analysis on this data, additional data analysis techniques are required. However, while further work is needed to get quantitative comparisons in this case, the fact that we are able to detect individual particle transits shows promise for further exploration.

\section{Conclusion}

We presented data on a new technique for cavity-enhanced aerosol detection using broadband light sources.  Using a superluminescent diode (SLD), we found that particles traversing the cavity mode created distinct peaks, but also that a background 'shoulder' was evident which was not explained by particles above 300~nm in size.  Comparing the number of fitted peaks to the number of particles measured by the Particles Plus counter shows that, while there are some disparities, the disparity trend is not fully explained by the shoulder.  Therefore, we conjecture that the shoulder is due to particles below 300~nm in size, marking that the cavity-enhanced particle counting method is more sensitive than traditional scattering methods.  Additionally, preliminary tests with LED-based versions of this setup suggest similar results, though further noise reduction and peak detection methods must be done.  Because of the difference in peak prominences and the shoulder shape between the SLD and LED experiments, we can conclude that a spatially-coherent light source (SLD) is more sensitive than a spatially-incoherent source (LED) as spatially-coherent sources can match to the cavity mode.  However, we can also conclude that LED-based detectors are feasible with enough noise reduction.

\subsection{Further Work}

We can further miniaturize the detector by increasing the reflectivity of the cavity mirrors to improve finesse and by using microfabrication techniques to fabricate custom light sources and photodiodes for the detector.  Additional study of the shoulder (Figure \ref{fig:backparticles}) is needed to fully characterize the sizes of particles detected, which could lead to an application in ultrafine particle detection.  We are considering new methods of analyzing the data generated by the LED experiment, such as calibrating the detector using mono-dispersed aerosols to more precisely assign peak parameters to specific particle sizes.  By doing so, a training data set could then be established for a machine-learning approach.



\begin{backmatter}
\bmsection{Funding}
Air Force Research Laboratory (FA9453-21-2-0064).

\bmsection{Acknowledgments}
This material is based on research sponsored by Air Force Research Laboratory (AFRL) under agreement number FA9453-21-2-0064. The U.S. Government is authorized to reproduce and distribute reprints for Governmental purposes notwithstanding any copyright notation thereon. The views and conclusions contained herein are those of the authors and should not be interpreted as necessarily representing the official policies or endorsements, either expressed or implied, of Air Force Research Laboratory (AFRL) and or the U.S. Government.

\bmsection{Disclosures}
The authors declare no conflicts of interest.



\bmsection{Data availability} Data underlying the results presented in this paper are not publicly available at this time but may be obtained from the authors upon reasonable request.

\bmsection{Supplemental document}
Link to supplement document

\end{backmatter}

\bibliography{references}






\end{document}